\documentstyle[aps,multicol,epsf]{revtex}
\newcommand{\be}{\begin{equation}}
\newcommand{\ee}{\end{equation}}
\begin{document}
\draft
\title{Nonequilibrium phase transition by directed Potts particles\\} 
\author{B.~Kahng and S.~Park\\}
\address{
Department of Physics and Center for Advanced Materials 
and Devices, Konkuk University, Seoul 143-701, Korea \\}
\maketitle
\thispagestyle{empty}

\begin{abstract}
We introduce an interface model with $q$-fold symmetry to 
study the nonequilibrium phase transition (NPT) from an active 
to an inactive state at the bottom layer.  
In the model, $q$ different species of particles are deposited 
or are evaporated according to a dynamic rule, which includes 
the interaction between neighboring particles within the same layer. 
The NPT is classified according to the number of species $q$. 
For $q=1$ and $2$, the NPT is characterized by directed percolation, 
and the directed Ising class, respectively. 
For $q \ge 3$, the NPT occurs at finite critical probability 
$p_c$, and appears to be independent of $q$; the $q=\infty$ case 
is related to the Edwards-Wilkinson interface dynamics. 
\end{abstract}


\begin{multicols}{2}
\narrowtext
Recently, the problems of phase transitions in nonequilibrium 
systems have attracted considerable interest in the physical 
literature\cite{review}. 
For example, the nonequilibrium phase transition (NPT) 
from an active state to an inactive state becomes one of 
the central issues in the field of nonequilibrium 
dynamics\cite{dickman}. 
It was shown that the number of equivalent inactive 
(or absorbing) states characterizes universality classes 
for the NPT \cite{park}. 
For example, NPT, occurring in the monomer-dimer model 
for the catalytic oxidation of CO \cite{zgb}, 
the contact process \cite{cp}, $etc$, 
has one absorbing state, and belongs to the directed 
percolation (DP) universality class \cite{dp}. 
Most of dynamic problems exhibiting the NPT with absorbing state, 
belong to this universality class. 
Meanwhile, there are a few exceptions: when there exist two 
absorbing states in the dynamic process, the critical behavior 
near the threshold of NPT is distinctive from that of the DP class, and 
belongs to the directed Ising (DI) class \cite{park}, 
equivalent to the class of parity-conserving branching-annihilating 
random walks \cite{baw}. 
The stochastic models in the DI class include, for example, 
the probabilistic cellular automata model\cite{grass}, 
the interacting monomer-dimer model\cite{kim}, 
the modified Domany-Kinzel model\cite{hinrichsen}. 
On the other hand, the scenario that the NPT can be classsified according 
to the number of absorbing states is not complete since 
the NPT with more than two absorbing states is not known yet. 
In other words, the NPT in the directed $q$-state Potts (DPotts) class 
for $q \ge 3$ is not discovered yet. 
One reason may be from that for $q \ge 3$,  
active sites (kinks) generated at the boundary between different 
states domains appear more often than in the DI case. 
Thus, the absorbing state could not be reached for finite
control parameter. Accordingly, it would be interesting 
to search for the NPT belonging to the DPotts class for $q \ge 3$. 

Recently, the NPT has been considered in association with the roughening 
transition (RT) from a smooth to a rough phase. 
The NPT behavior appears at a particular reference height of the 
interface. For example, for the monomer deposition-evaporation 
(m-DE) model introduced by Alon $et$ $al$ \cite{alon}, where 
evaporation of particle is not allowed on terrace, 
the reference height is the spontaneously selected bottom layer. 
The site where the interface touches the reference height, 
called vacant site, corresponds to the active site of the NPT.   
In the active phase of the NPT, the interface fluctuates close to the 
reference height, being in a binded state, in which the interface 
is smooth. On the other hand, in the inactive phase, 
the interface is detached from the reference height, being 
in a unbinded state, and the interface is rough. 
Accordingly, RT accompanied by the binding-unbinding 
transition of the interface is related to the NPT at the reference height, 
which is characterized according to the number of symmetric states in 
the dynamic rule\cite{parks}. For the interface models 
with two-fold symmetry in their dynamic rule,  
which are the extensions of the m-DE model by assigning 
a couple of species to particles for one model \cite{parks}, called the 
two-species model, and by enlarging the size of particles 
for the other model \cite{odor}, called the dimer DE (d-DE) model,  
the dynamics at the reference height near the threshold 
of RT behaves similarly to the DI dynamics.

Both models contain the suppression effect against generating 
active sites, and the critical threshold of RT is considerably 
lower compared with the monolayer version. 
For the two-species model, the critical threshold is $p_c\approx 0.4480$ 
for the interface version, but $p_c\approx 0.7485$ 
for the monolayer version. 
In this Letter, we introduce an interface model with 
arbitrary $q$-fold symmetry, called the $q$-species model, 
and examine the NPT at the reference height for the cases 
$q=3,4,5$ and $\infty$. The result is compared with the previous one 
for $q=1$ and 2. In addition, we study RT for each $q$.  
Interestingly, it is found that the NPT at the reference 
height $occurs$ at finite deposition-probability for all cases, 
and their characteristics for $q \ge 3$ are different from the cases 
$q=1$ and 2, but appear to be independent of $q$ as long as 
$q \ge 3$. In particular, the interface model for the case 
$q=\infty$ is reduced to the restricted solid-on-solid (RSOS) model 
with deposition and evaporation processes. In this case, 
since RT occurs when the probabilities of deposition and 
evaporation are equal, the interface dynamics belongs to 
the Edwards-Wilkinson (EW) interface dynamics \cite{ew}. 
Accordingly, it is found that the NPT of the $q$-species model 
for $q\ge 3$ in one dimension is related to the EW class. 
  
In the $q$-species model, $q$ different species particles
are deposited or evaporated on one dimensional substrate 
with periodic boundary condition. A site is first selected at random 
at which either deposition or evaporation of a particle is attempted 
with probability $p$ and $1-p$, respectively. 
In the attempt of deposition,  
one species is selected among the $q$ species with equal 
weight $1/q$. Deposition or evaporation are realized 
under two conditions described below. First, the RSOS condition 
is imposed such that the height difference between nearest 
neighboring columns does not exceed one. 
Thus the attempt is realized as long as the RSOS 
condition is satisfied even after deposition or evaporation event.  
Secondly, the interaction between neighboring particles within 
the same layer is considered: 
When a particle of a certain species (e.g. A-species) is deposited 
on a hollow between particles, the deposition is not allowed when both 
of the neighboring particle on each side are of a common species 
(B-species), different from the dropping one (A-species). 
Meanwhile, a particle of a certain species (e.g. A-species) is not 
allowed for evaporation when it is sandwiched between particles of 
the same species (A-species) within the same layer.   
However, a particle can deposit or evaporate when two neighboring 
particles on each sides are of different species from one another,  
or one (or both) of the neighboring sites is (are) vacant.
As $q \rightarrow \infty$, the probability of forming three particles 
in a row with the same species is zero. 
Thus, the secondary restriction is meaningless. 
Hence, the model is reduced to a random deposition-evaporation 
model under the RSOS condition. On the other hand, the initial 
substrate is flat where its height is referred as $h=0$. 
The dynamics proceeds only for $h \ge 0$, so that evaporation 
of particles at $h=0$ is not allowed.  

Monte Carlo simulations are performed by varying the deposition 
probability $p$ and system size $L$ for the cases $q=3,4,5$ and 
$\infty$. We first discuss the dynamics occurring at 
the reference height, $h=0$. 
We consider the density of the vacant site $\rho(p,t)$ at the 
reference height with varying time $t$ at a certain 
deposition probability $p$. 
When $p$ is small, particles form small-sized 
islands which disappear after their short lifetime 
and the growth velocity of interface is zero. 
The density $\rho(p,t)$ saturates at finite value 
as $t \rightarrow \infty$, as shown in Fig.1. 
As $p$ increases, deposition increases and islands grow, 
until, above a critical value $p_c$, islands merge and fill 
new layers completely, giving the interface a finite growth 
velocity. Accordingly, RT occurs at $p_c$, and 
the NPT at the reference height occurs as well. 
The critical probability $p_c$, estimated for each $q$, 
is listed in Table 1. 
In particular, for $q=\infty$, the critical probability 
$p_c$ is determined as $p_c=0.5$, when the probabilities 
of deposition and evaporation are equal. 
For $p < p_c$, the saturated value $\tilde{\rho}(p)$ behaves as 
\be
{\tilde{\rho}}(p) \sim (p_c-p)^{\beta}, 
\ee
with the order parameter exponent $\beta$. 
We estimated the exponent $\beta$ from the data obtained 
from system size $L=500$, which is listed in Table 1.  
The numerical values for $q \ge 3$ do not seem to be close 
to each other, which makes it hard to conclude that 
the cases of $q \ge 3$ belong to the same universality class. 
However, the exponent $\beta$ is hard to be measured precisely, 
because it is extremely sensitive to the estimated value $p_c$. 
Accordingly, the numerical value of $\beta$ even for $q=2$ is 
rather broadly ranged as can be noticed in Ref.\cite{hinrichsen}.  
On the other hand, for $p > p_c$, the density decreases to 
zero exponentially in the long time limit and at the critical 
threshold $p_c$, it decays algebraically as 
\be 
\rho(p_c,t) \sim t^{-\beta/\nu_{\parallel}}. 
\ee
The power, $\beta/\nu_{\parallel}$, is measured 
in Fig.1 and is tabulated in Table 1. 
The values are reasonably close to one another for $q \ge 3$, 
but considerably larger than the values for $q=1$ and 2. 
This result suggests that the number of species is 
unimportant for $q \ge q_c \equiv 3$. 
On the other hand, we need to discuss the relation between the 
vacant sites and the kinks which are the domain boundaries 
between different species, because the kinks are indeed active 
sites in the directed Potts dynamics. 
We measure the kink density $\rho_K(p_c,t)$ by counting the sites 
with height equal to one and whose neighbors are different species,
which behaves similarly to $\rho(p_c,t)$, as shown in Fig.2. 
Thus we confirm that the vacant site density is 
equivalent to the active site density in the $q$-species model. 
Finally, we simulate the monolayer version of the $q$-species 
model, and find that the system does not exhibit NPT and 
is always in the active state for finite $p$. 
A similar behavior was found in the three-species monomer-monomer 
reaction model introduced by Bassler and Browne \cite{bassler}, 
in which the system is always in a reactive phase when the absorption 
rates for each species are identical.
 
The density $\rho(p,t)$ can be thought as 
the return probability $P_0(p,t)$ of interface height 
$h(x,t)$ to its initial height $h(x,t)=0$ after passing 
time $t$, averaged over the substrate position $x$. 
The subscript $0$ means that the time is measured from $t=0$. 
In general, $P_0(p,t)$ is different from the first return 
probability $F_0(p,t)$ of the interface height to its initial height, 
which decays as $F_0(p,t)\sim t^{-\theta}$ where $\theta$ 
is called the persistent exponent \cite{krug}. 
For the EW interface under the given boundary condition at $h=0$, 
we obtain that $F_0(p_c,t) \sim t^{-0.75}$, similar to 
$P_0(p_c,t)$. 
Hence, the power $\beta/\nu_{\parallel}$ for $q=\infty$ is related 
to the persistent exponent for the EW interface. 

Next, let us consider the size dependence of $\rho(p_c,t)$. 
To do so, we study the averaged density $\rho^{(s)}(p_c,t)$ over the 
active runs which contain at least one vacant site at the reference 
height. Since $\rho(p_c,t)$ decays as Eq.(2) at $p_c$,  
the vacant sites disappear completely as $t\rightarrow \infty$. 
There exists a characteristic time $\tau$ such that every run is 
active up to the time $\tau$, while for $t > \tau$, 
it is active with probability less than one and 
proportional to $\rho(p_c,t)$, leading to the saturation of 
$\rho^{(s)}(p_c,t)$, as shown in Fig.3. 
This fact implies that for $t > \tau$, when vacant sites exist, 
their number should be of order of unity. Once they are occupied, 
the sample falls into the inactive state, because other sites are 
covered by particles on upper layer. 
Hence, the density follows $\rho(p_c,\tau)\approx 1/L$. 
The characteristic time $\tau$ depends on system size $L$ 
as $\tau \sim L^{z}$ with the dynamic exponent 
$z=\nu_{\parallel}/\nu_{\perp}$. 
According to the scaling theory, the saturated value, 
${\tilde \rho}^{(s)}(p_c)$ is written as 
\be 
{\tilde \rho}^{(s)}(p_c) \sim L^{-\beta/\nu_{\perp}}.    
\ee 
The power $\beta/\nu_{\perp}$ is found to be $\approx 1$, 
as predicted. Therefore, the dynamic exponent 
$z$ is obtained to be between 1.32 and 1.41, which 
deviates from the value $z_{EW}=2$ for the interface dynamics. 
The origin of this discrepancy may be that the dynamics 
at the reference level is affected by the boundary at $h=0$. 

Let us consider the dynamics of interface above the reference 
height. We examine the interface fluctuation width, 
$W^2(L,t)={1\over L}\sum_x h^2(x,t)-\big({1\over L}
\sum_x h(x,t) \big)^2$ at $p_c$. 
Contrary to the cases $q=1$ and 2, $W^2$ exhibits 
the power-law behavior as shown in Fig.4,  
\be 
W^2(L,t) \sim \cases{
t^{2\zeta}, 
& for $t \ll L^{\chi/\zeta}$, \cr 
L^{2\chi}, 
& for $t \gg L^{\chi/\zeta},$ \cr} 
\ee
where $\zeta$ and $\chi$ are the growth and the roughness 
exponents, respectively. 
The exponents $\zeta$ and $\chi$ for $q \ge 3$
are obtained numerically as  
$\zeta \approx 0.22, 0.23, 0.23$ and $0.24$, and 
$\chi \approx 0.46, 0.48, 0.46$ and $0.48$ 
for $q=3,4,5$ and $\infty$, respectively. 
These values are close to the EW values, 
$\zeta=1/4$ and $\chi=1/2$. 
We also examine the height-height correlation function, 
$C^2(r)=<{(h(r)-h(0))^2}>\sim r^{2\chi'}$ in the long 
time limit. The exponent $\chi'$ is consistent 
with $\chi$. 
For $p > p_c$, the scaling of the interface width belongs 
to the Kardar-Parisi-Zhang universality class 
\cite{kpz}. 

We also generalize the d-DE model into the trimer case, 
called the t-DE model, where a trimer is deposited with probability 
$p$ or evaporated with probability $1-p$ according to a rule similar 
to the d-DE model. Note that the t-DE model is 
different from the model introduced by Stinchcombe $et.al.$ 
\cite{stinchcombe} in the two aspects: 
the former (latter) model is an interface (monolayer) model, 
and evaporation on terrace is prohibited (allowed).    
As shown in Fig.5, there exists a critical deposition 
probability $p_c$ such that for $p < p_c$, 
the density of the vacant sites at the bottom layer 
is saturated, whereas for $p \ge p_c$, it decays 
algebraically. Note that the behavior 
of the exponential-type decay for $p > p_c$ does not 
appear in the t-DE model. 
The exponent, $\beta/\nu_{\parallel}$ is obtained to be 
$\approx 0.38$, which is almost half of the value for 
the 3-species model, which is also the case for $q=2$. 
On the other hand, the dynamic exponent $z\approx 2.47$ 
is obtained by measuring the characteristic time 
for finite-size cutoff.  
This value is inconsistent with the one for 
the $q$-species model. 
Accordingly, further study for the dynamic exponent 
$z$ is required. Detailed numerical results for the t-DE 
model and its generalization into arbitrary $q$-mer case 
will be published elsewhere \cite{kahng}.

In summary, we have introduced the $q$-species model, 
which is an interface model exhibiting RT accompanied 
by the binding-unbinding transition with respect to the 
reference height. The NPT occurring at the reference height 
has been examined numerically for $q=3,4,5$ and $\infty$. 
We found that the NPT occurs at finite $p_c$ for 
all the cases of $q$, which is remarkable, because the NPT 
for $q \ge 3$ has never been found before. 
For $q \ge q_c=3$, the number of species is unimportant, 
and the NPT independent of $q$ is related to the EW interface 
dynamics via the return probability of interface to its 
initial height.   
We measured the critical exponents for NPT and RT.
We also considered the t-DE model with three-state 
symmetry, and compared the result with the three-state model.
 
The authors wish to thank H.~Park and J.D.~Noh 
for helpful discussions. This work was supported 
in part by the Korea Research Foundation 
(98-001-D00280 \& 98-015-D00090).\\


\begin{figure}
\centerline{\epsfxsize=7.5cm \epsfbox{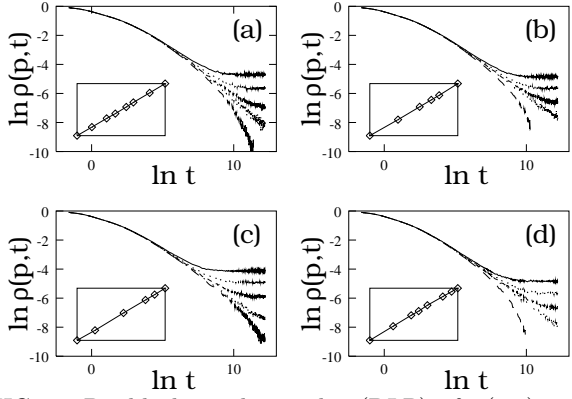}}
\caption{Double logarithmic plot (DLP) of $\rho(p,t)$ 
versus time $t$ (a) for $q=3$ at $p=0.4904$, 
$0.4910$, $0.4913$, $0.4914(=p_c)$ and $0.4917$ from the top.
(b) for $q=4$ at $p=0.5020$, $0.5024$, $0.5026$, 
$0.5027(=p_c)$, and $0.5031$ from the top. 
(c) for $q=5$ at $p=0.5051$, $0.5059$, $0.5063$, $0.5065(=p_c)$, 
and $0.5066$ from the top. 
(d) for $q=\infty$ at $p=0.4993$, $0.4997$, $0.4999$, $0.5000(=p_c)$, 
and $0.5005$ from the top. 
All data are obtained for $L=500$, averaged over 1000 realizations.
Insets: DLP of $\tilde{\rho}(p)$ versus $\epsilon$ for each $q$.
The solid line is a guideline to the eye.}
\label{fig1}
\end{figure}

\begin{figure}
\centerline{\epsfxsize=6.0cm \epsfbox{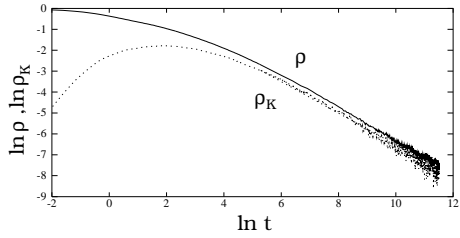}}
\caption{DLP of $\rho(p_c,t)$ (top) and the kink density 
$\rho_K(p_c,t)$ versus time $t$ at $p_c$. The numerical 
results are obtained for $L=1000$, averaged over 1000 
realizations.}
\label{fig2}
\end{figure}

\begin{figure}
\centerline{\epsfxsize=7.5cm \epsfbox{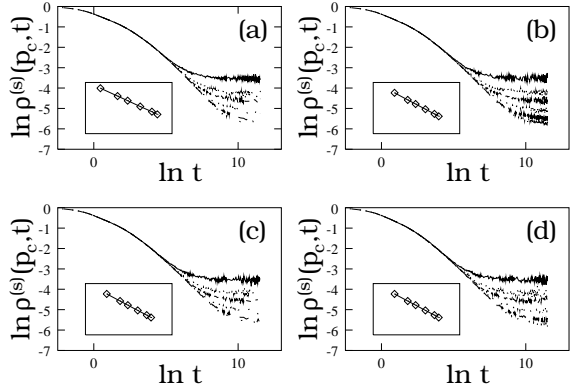}}
\caption{DLP of $\rho^{(s)}(p_c,t)$ 
versus time $t$ at $p_c$ for different 
system sizes $L=100$, $200$, $300$, $500$, $800$ and $1000$ from 
the top for (a) $q=3$, (b) $q=4$, (c) $q=5$, and (d) $q=\infty$.
The data are averaged over 1000 realizations.
Insets: DLP of ${\tilde \rho^{(s)}}(p_c)$ versus $L$ for 
each $q$. The solid line is a guideline to the eye.}
\label{fig3}
\end{figure}

\begin{figure}
\centerline{\epsfxsize=7.5cm \epsfbox{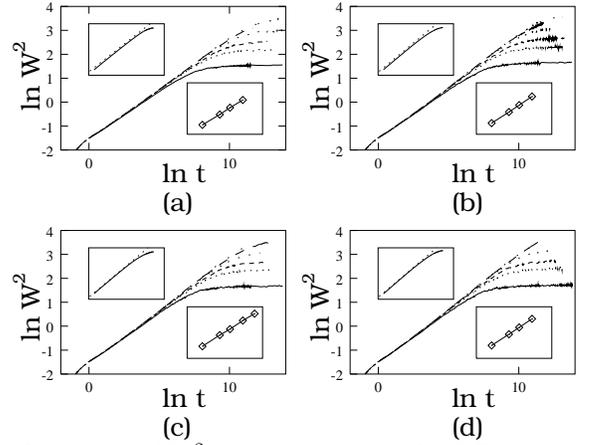}}
\caption{DLP of $W^2$ versus time $t$ at $p_c$ for different 
system sizes $L=100, 200, 300, 500, 800$ and $1000$ 
from the bottom for (a) $q=3$, (b) $q=4$, (c) $q=5$, and 
(d) $q=\infty$. Insets-left: DLP of $C^2(r)$ versus distance 
$r$ for $L=500$ for each $q$. The dashed line is a guideline 
to the eye. 
Insets-right: DLP of the saturated value of $W^2$ versus $L$ 
for each $q$. The solid line is a guideline to the eye. 
All the data are averaged over 1000 realizations} 
\label{fig4}
\end{figure}

\begin{figure}
\centerline{\epsfxsize=6.0cm \epsfbox{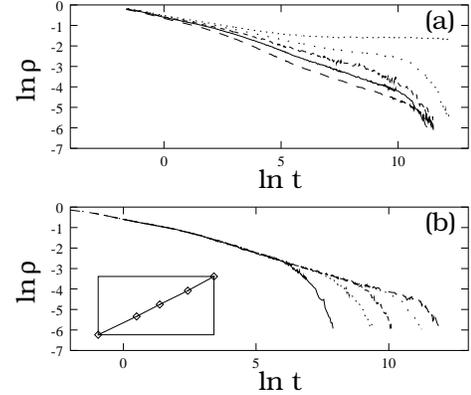}}
\caption{(a) DLP of $\rho(p,t)$ versus time $t$ 
for the t-DE model at $p=0.32, 0.34, 0.36$, $0.38(=p_c)$, 
and $0.42$ from the top. The data are obtained 
from system size $L=500$, averaged over 500 realizations.
(b) DLP of $\rho(p_c,t)$ versus time $t$ for 
various system sizes $L=99,198,300,498$ and 798 from 
the left. The data are averaged over 500 realizations. 
Inset: DLP of the characteristic time $\tau$ 
versus system size $L$. The solid line is a guideline 
to the eye.}
\label{fig5}
\end{figure}

\begin{table}
\caption{The estimated values of $p_c$, 
$\beta$, $\beta/\nu_{\parallel}$, $\beta/\nu_{\perp}$, 
and $z$ for various $q$. The data for the cases of 
$q=1$ and $q=2$ are quoted from Refs.12 and 13, 
respectively.}  
\begin{center}
\begin{tabular}{ccccccccccc}
 & $q$ & $p_c$ & $\beta$ & $\beta/\nu_{\parallel}$ & $\beta/\nu_{\perp}$ & 
$z$ &\\
\hline
 &1 & 0.189$\phantom{x}$ & 0.28 & 0.16 & 0.25 & 1.56 &\\
 &2 & 0.4480 & 0.88 & 0.59  & 0.97 & 1.64 &\\
 &3 & 0.4914 & 1.01(3) & 0.75(1) & 0.99(1) & 1.32 &\\
 &4 & 0.5027 & 0.94(2) & 0.73(1) & 0.99(1) & 1.36 &\\
 &5 & 0.5065 & 0.91(3) & 0.70(1) & 0.99(1) & 1.41 &\\
 & $\infty$ & 0.5000 & 0.95(3) & 0.75(1) & 0.99(1) & 1.32   
\end{tabular}
\end{center}
\end{table}
\end{multicols}
\vfil\eject

\begin{thebibliography}{99}
\vspace{-1.0cm}
\bibitem{review} For a review, see, V. Privman, 
{\it Nonequilibrium Statistical Mechanics in One Dimension} 
(Cambridge University Press, Cambridge, 1996).  
\bibitem{dickman} J. Marro and R. Dickman,  
{\it Nonequilibrium Phase Transitions} 
(Cambridge University Press, Cambridge, 1997).  
\bibitem{park} H. Park and H. Park, {\rm Physica A} {\bf 221,} 97 (1995).
\bibitem{zgb} R.M. Ziff, E. Gulari and Y. Barshad, 
{\rm Phys. Rev. Lett.} {\bf 56,} 2553 (1985). 
\bibitem{cp} T.E. Harris, {\rm Ann. Prob.} {\bf 2}, 969 (1974); 
T.M. Liggett, {\it Interacting Particle Systems} 
(Spinger-Verlag, New York,1985)   
\bibitem{dp} G. Deutscher, R. Zallen and J. Adler, 
{\it Percolation Structures and Processes} Ann. Isr. Phys. Soc.5 
(Adam Hilger, Bristol, 1983).
\bibitem{baw} H. Takayashu abd A. Y. Tretyakov, {\rm Phys. Rev. Lett.} 
{\bf 68,} 3060 (1992).
\bibitem{grass} P. Grassberger, F. Krause and T. von der Twer, 
{\rm J. Phys. A} {\bf 17}, L105 (1984). 
\bibitem{kim} M.H. Kim and H. Park, {\rm Phys. Rev. Lett.} {\bf 73,} 2579 
(1994).
\bibitem{hinrichsen} H. Hinrichsen, {\rm Phys. Rev. E} {\bf 55,} 219 (1997).
\bibitem{alon} U. Alon, M.R. Evans, H. Hinrichsen and D. Mukamel, 
{\rm Phys. Rev. Lett.} {\bf 76,} 2746 (1996); {\rm Phys. Rev. E} 
{\bf 57}, 4997 (1998).
\bibitem{parks} S. Park and B. Kahng (cond-mat/9807193). 
\bibitem{odor} H. Hinrichsen and G. \'Odor, {\rm Phys. Rev. Lett.} 
{\bf 82}, 1205 (1999). 
\bibitem{ew} S.F. Edwards and D.R. Wilkinson, 
{\rm Proc. R. Soc. Lond. A} {\bf 381,} 17 (1982). 
\bibitem{bassler} K.E. Bassler and D.A. Browne, {\rm Phys. Rev. Lett.} 
{\bf 77}, 4094 (1996); {\rm Phys. Rev. E} {\bf 55,} 5225 (1997).  
\bibitem{krug} J. Krug, H. Kallabis, S.N. Majumdar, S.J. Cornell, 
A.J. Bray, and C. Sire, {\rm Phys. Rev. E} {\bf 56,} 2702 (1997).
\bibitem{kpz} M. Kardar, G. Parisi and Y.C. Zhang, {\rm Phys. Rev. Lett.} 
{\bf 56,} 889 (1986). 
\bibitem{stinchcombe} R.B. Stinchcombe and M.D. Grynberg, and 
M. Barma, {\rm Phys. Rev. E} {\bf 47,} 4018 (1993).
\bibitem{kahng} B. Kahng and S. Park (unpublished).
\end{thebibliography}
\end{document}